# PANDA: Predicting the change in proteins binding affinity upon mutations using sequence information

Wajid Arshad Abbasi[1*], Syed Ali Abbas[1], Saiqa Andleeb[2]

[1]Computaional Biology and Data Analysis Lab., Department of Computer Sciences & Information Technology, King Abdullah Campus,  University of Azad Jammu & Kashmir, Muzaffarabad, AJ&K, 13100, Pakistan.

[2]Biotechnology Lab., Department of Zoology, King Abdullah Campus, University of Azad Jammu & Kashmir, Muzaffarabad, AJ&K, 13100, Pakistan.

*Corresponding author's emails: wajidarshad@gmail.com; wajidarshad@ajku.edu.pk

## Abstract

Accurately determining a change in protein binding affinity upon mutations is important for the discovery and design of novel therapeutics and to assist mutagenesis studies. Determination of change in binding affinity upon mutations requires sophisticated, expensive, and time-consuming wet-lab experiments that can be aided with computational methods. Most of the computational prediction techniques require protein structures that limit their applicability to protein complexes with known structures. In this work, we explore the sequence-based prediction of change in protein binding affinity upon mutation. We have used protein sequence information instead of protein structures along with machine learning techniques to accurately predict the change in protein binding affinity upon mutation. Our proposed sequence-based novel change in protein binding affinity predictor called PANDA gives better accuracy than existing methods over the same validation set as well as on an external independent test dataset. On an external test dataset, our proposed method gives a maximum Pearson correlation coefficient of 0.52 in comparison to the state-of-the-art existing protein structure-based method called MutaBind which gives a maximum Pearson correlation coefficient of 0.59. Our proposed protein sequence-based method, to predict a change in binding affinity upon mutations, has wide applicability and comparable performance in comparison to existing protein structure-based methods. A cloud-based webserver implementation of PANDA and its python code is available at https://sites.google.com/view/wajidarshad/software and https://github.com/wajidarshad/panda.

## Introduction

Protein interactions play a crucial role in cellular biology through metabolic pathway regulation and by maintaining homeostasis (Du et al., 2016; Vincenzi and Leone, 1969). These interactions have associated binding affinity which determines the stability and specificity of proteins upon complex formation (Du et al., 2016). Therefore, to completely decipher the cellular processes not only requires a thorough understanding of protein-protein interactions mechanism but also the quantitative knowledge of associated binding affinity. Mutations in protein sequences modulate or often disrupt protein interactions by changing associated binding affinity which may forbid normal protein function and cause disease (Cargill et al., 1999; Li et al., 2016; Siebenmorgen and Zacharias, 2019). Therefore, measuring the change in binding affinity upon mutation is important to assess the mutational effects while designing new synthetic protein-protein complexes with the desired function and novel drugs. The change in binding affinity is often measured in terms of change in the Gibbs free energy upon interaction ($\Delta\Delta G$) (Bitencourt-Ferreira and de Azevedo, 2018). Accurate measurement of change in binding affinity upon mutation is possible through different wet-lab experimental techniques such as site-directed mutagenesis, Isothermal Titration Calorimetry (ITC), Surface Plasmon Resonance (SPR), and Fluorescence Polarization (FP) (Jönsson et al., 1991; Kastritis and Bonvin, 2013; Li et al., 2016; Perozzo et al., 2004; Weber, 1952). However, these experimental techniques are expensive and time-consuming which may prohibit their large-scale use. Therefore, the expansion of reliable *in-silico* approaches to accurately predict the effect of mutations on change in binding affinity of protein complexes is immensely required to assist the wet-lab experiments.

Among *in-silico* approaches to accurately predict the effect of mutations on change in binding affinity, machine learning is preferable because of its implicit treatment of unknown factors involved in protein interaction and its ability to learn a flexible data-driven function (Ain et al., 2015; Silva et al., 2020; Xavier et al., 2016). A bunch of machine learning based methods has been proposed in the literature to predict the change in binding free energy ($\Delta\Delta G$) upon mutations (Berliner et al., 2014; Brender and Zhang, 2015; Chen et al., 2019; Geng et al., 2019; Li et al., 2016; Pahari et al., 2020; Pires et al., 2014; Rodrigues et al., 2019; Witvliet et al., 2016; Zhang et al., 2020). Among these proposed methods, a few have been reported reasonably accurate on predicting change in binding affinity on mutations (Geng et al., 2019; Li et al., 2016; Pahari et al., 2020; Rodrigues et al., 2019; Witvliet et al., 2016; Zhang et al., 2020). However, these methods cannot be applied on a larger scale because of their dependence on 3D structures of the protein complexes which are usually not readily available. This problem of unavailability of 3D structures of protein complexes further gets worse while studying mutational effects where 3D structures are required at each mutation. Therefore, there is a need to design alternate computational methods with reasonable accuracy based on easily available sources.

Among sequence-based change in protein binding affinity prediction models, the model involving deep learning proposed by Chen *et al*. is the state-of-the-art predictor (Chen et al., 2019). Chen *et al*. reported a higher accuracy with a correlation coefficient of 0.873 using 10-fold cross-validation and SKEMPI V2.0 dataset (Chen et al., 2019; Moal and Fernández-Recio, 2012). However, the cross-validation scheme adopted by Chen *et al*., may not conform to the underlying problem as

SKEMPI V2.0 dataset involves more than one mutant proteins of a single protein complex (Abbasi and Minhas, 2016; Chen et al., 2019; Moal and Fernández-Recio, 2012). Moreover, protein binding affinity prediction models proposed by Chen *et al*. had not been evaluated using any external validation datasets, and also this study did not provide an interface to perform such a validation (Chen et al., 2019). Therefore, there is a need to revisit sequence-based change in binding affinity prediction upon mutations and to develop novel predictors that can be used in a practical setting. In this study, we present a state-of-the-art machine learning based predictor called PANDA (Predict chAnge iN binDing Affinity) to predict a change in protein binding affinity upon mutations. Our proposed predictor uses protein sequence information only and gives reasonable accuracy with its wide applicability.

## Methods

**Dataset for experimentation and its preprocessing**

We have used a dataset of protein sequences extracted from the 3D structures of protein complexes present in the SKEMPI V2.0 database (Moal and Fernández-Recio, 2012). This dataset contains the experimentally determined $\Delta\Delta G (kcal/mol)$ values for both wild type and mutated protein complexes (Jankauskaitė et al., 2019; Moal and Fernández-Recio, 2012). The dataset used in this study contains 1215 mutations from 81 complexes after removing complexes with mutations in both chains, by selecting only dimeric complexes and complexes with more than 10 recorded mutations. This set is very similar to the dataset used to develop the MutaBind (Li et al., 2016). We have also extracted a validation dataset from the SKEMPI V2.0 database with 129 mutations records in 19 different dimeric complexes. We have selected complexes in the validation dataset with less than 10 recorded mutations.

**Proposed Methodology**

We have developed a regression model based on protein sequence information called PANDA (Predict chAnge iN binDing Affinity) to predict a change in protein binding affinity upon mutation. To develop PANDA, we have used different regression techniques, evaluation protocols, and sequence-based feature extraction techniques. The methodology adopted for the development of PANDA is as follows.

**Sequence-based features**

To develop a machine learning based prediction model, we require to represent each train and test example in our dataset with an appropriate feature vector. Therefore, we have represented each complex in our dataset through a feature representation obtained from individual chains in the ligand (l) and receptor (r) of each complex before and after mutations. We have used a number of explicit features to model sequence-based attributes of protein complexes. We discuss the sequence-based features used in this study below.

**k-mer composition (k-mer)**

The k-mer composition, obtained by counting the occurrences of individual k-mer in a protein sequence, is a widely used descriptor of a protein sequence (Leslie et al., 2002). We used this feature representation to capture the composition of a protein sequence. This feature representation gives a feature vector $\phi_{k-mer}(s)$ of a given sequence $s$ such that the $\phi_{k-mer}(s)_k$ contains the number of times $k-mer$ occurs in $s$. For k-mers of size 1 (1-mer), 2 (2-mer) and 3 (3-mer), this yields a 20, 400 and 8000-dimensional feature representation $\phi_{k-mer}(s)$ of each protein chain $s$, respectively. This feature representation has successfully been used to predict protein interactions, binding sites and prion activity (Abbasi et al., 2018; Leslie et al., 2002; Minhas et al., 2017; Minhas and Ben-Hur, 2012).

**BLOSUM-62 features (Blosum)**

In order to represent the amino acid composition and at the same time to capture substitutions of physiochemically similar amino acids in a protein sequence, a protein sequence is converted into a 20-dimensional vector $\phi_{Blosum}(s)$ by averaging the columns from a BLOSUM substitution matrix corresponding to each amino acid in a given protein chain $s$. We have used a BLOSUM-62 substitution matrix to extract this feature representation (Eddy, 2004). This feature representation has already been used successfully in several related studies (Abbasi et al., 2018; Abbasi and Minhas, 2016; Aumentado-Armstrong et al., 2015; Westen et al., 2013; Zaki et al., 2009).

**Propy features (propy)**

In order to capture the biophysical properties of amino acids and sequence-derived structural features of a given protein sequence, we have used a feature extraction package called propy (Cao et al., 2013). It gives a 1,537-dimensional feature representation $\phi_{propy}(s)$ of a given protein chain $s$. This representation includes pseudo-amino acid compositions (PseAAC), autocorrelation descriptors, sequence-order-coupling number, quasi-sequence-order descriptors, amino acid composition, transition and the distribution of various structural and physicochemical properties (Li et al., 2006; Limongelli et al., 2015).

**Structure-based features (MutaBind-terms)**

We have also used structure-based features of the protein complexes involved in our dataset in order to get a direct comparison with sequence-based models. These structure-based features have been used by Li *et al*. and we have obtained these features from the supplementary material (Li et al., 2016). These structure-based features include different terms such as Van der Waals interaction energies, differences between polar solvation energies, the difference between unfolding free energies, solvent accessible surface areas, conservation score, and proline's cyclic structure (Li et al., 2016).

**Complex level feature representation**

In our machine learning setting, a complex $c$ is represented by the tuple $c = \big((l,r),(l',r'),y\big)$, where $(l,r)$ and $(l',r')$ are the pairs of ligand and receptor proteins in the wild and mutated complex, respectively, and $y$ is the corresponding change in binding affinity value. To generate the complex level feature representation $\psi(c)$, we simply subtract the feature representations of respective wild $(c_w)$ and mutated $(c_m)$ type complexes as $\psi(c) = [\psi(c_w) - \psi(c_m)]$ where, $\psi(c_w) = \begin{bmatrix}\phi(l)\\\phi(r)\end{bmatrix}$ and $\psi(c_m) = \begin{bmatrix}\phi(l')\\\phi(r')\end{bmatrix}$.

## Regression models

In this study, we have presented the change in binding affinity prediction upon mutations as a regression and classification problem. In machine learning based change in affinity prediction, we have a dataset consisting of $N$ examples of the form $(\boldsymbol{\psi}(c_i), y_i)$, where $i = 1 \ldots N$. In this representation, $\boldsymbol{\psi}(c_i)$ is a feature representation of a complex $c_i$ with a known change in binding affinity $y_i$ upon mutation. Our objective in machine learning based regression and classification is to train a model $f(c)$ that can predict the change in binding affinity of the complex $c$ upon mutations in its constituent chains. For the classification of increase or decrease in binding free energy of a complex $c_i$ upon mutation, we have used predicted scores of the trained regression model along with $\boldsymbol{y_i} \in \{+\mathbf{1}, -\mathbf{1}\}$ is its associated label indicating whether $\boldsymbol{c_i}$ has an increase in binding free energy $(+\mathbf{1})$ or decrease $(-\mathbf{1})$. The learned regression function $f(\cdot)$ should generalize well over previously unseen complexes. We used the following regression techniques through Scikit-learn to get different regression models (Pedregosa et al., 2011). It is also important to note that the feature representations are normalized to have unit norm and standardized to zero mean and unit standard deviation before using them in the regression model.

## Support Vector Regression (SVR)

Support Vector Machines (SVMs) have effectively been used to solve different computational problems in bioinformatics (Cortes and Vapnik, 1995). Support Vector Regression (SVR) performs regression using $\varepsilon$-insensitive loss and, by controlling model complexity (Smola and Schölkopf, 2004). Training a SVR for protein binding affinity prediction involves optimizing the objective function given in Eq. (1) to learn a regression function $f(c) = \boldsymbol{w}^T \boldsymbol{\psi}(c) + b$.

$$min_{w,b} \frac{1}{2} \|\boldsymbol{w}\|^2 + C \sum_{i=1}^{N} (\xi_i^+ + \xi_i^-)$$

$$Such\ that\ for\ all\ i: \begin{cases} y_i - f(\boldsymbol{c}_i) \leq \varepsilon + \xi_i^+ \\ f(\boldsymbol{c}_i) - y_i \leq \varepsilon + \xi_i^- \\ \xi_i^+, \xi_i^- \geq 0 \end{cases} \quad (1)$$

Here, $\frac{1}{2}\|\boldsymbol{w}\|^2$ controls the margin, $\xi_i^+$ and $\xi_i^-$ capture the extent of margin violation for a given training example and $C$ is the penalty of such violations (Cortes and Vapnik, 1995). We used both linear and radial basis function (rbf) SVR in this study. The values of C, gamma, and epsilon were optimized during model selection. SVR has already been used for the same purpose in previous studies (Abbasi et al., 2017; Yugandhar and Gromiha, 2014).

## Random Forest Regression (RFR)

Random Forest Regression (RFR) is an ensemble of regression trees used for nonlinear regression (Breiman, 2001). Each regression tree in the RF is based on randomly sampled subsets of input features. We optimized RF with respect to the number of decision trees and a minimum number of samples required to split in this study. This regression technique has been used in many related studies (Abbasi et al., 2017; Ballester and Mitchell, 2010; Li et al., 2014; Moal et al., 2011).

### XGBoost regression (XGBR)

XGBoost is based on the boosting technique of an ensemble. It combines weak learners into a strong learner in an iterative fashion (Chen and Guestrin, 2016; Friedman, 2001). We have used default base learners of XGBoost that are tree ensembles. We have performed a model selection for XGBoost in terms of the number of boosting iterations, booster, subsample ratio, learning rate, and maximum depth using a grid search and xgboost 0.7 (Chen and Guestrin, 2016).

### Model validation, selection and performance assessment

To evaluate the performance of our trained regression models, we have used Leave One Complex Out (LOCO) and 10-fold cross-validation (CV) (Abbasi et al., 2018; Abbasi and Minhas, 2016). In LOCO, a regression model is developed with all the mutations in $(N - 1)$ complexes and tested on all the mutations of the left out complex. This process is repeated for all the $N$ complexes present in the dataset. Whereas, in 10-fold CV, we have shuffled mutations in all the complexes and then split them into 10 groups. Ten models have been trained and evaluated with each group given a chance to be held out as the test set. For the regression problem, we used Root Mean Squared Error $RMSE = \sqrt{\frac{1}{n}\sum_{i=1}^{N}(y_i - f(c_i))^2}$ , absolute error (AE), Pearson correlation coefficient $(P_r)$ and the Spearman correlation coefficient $(S_r)$ between the predicted $f(c_i)$ and actual $y_i$, as performance measures for model evaluation and performance assessment. To check the statistical significance of the results, we have also estimated the P-value of the correlation coefficient scores. Similarly, for the classification task, we have used the area under the ROC curve (ROC) and the area under the precision-recall curve (PR) as performance measures for model evaluation and performance assessment (Abbasi and Minhas, 2016; Davis and Goadrich, 2006). We used grid search over training data to find the optimal values of hyper-parameters of different regression models.

## Results and discussion

In this work, we have proposed a protein sequence-based machine learning based method to predict the change in protein binding affinity upon mutations. For this purpose, we have used various machine learning algorithms and different sequence-based features. In what follows we present results showing the prediction performance of our proposed method across two different cross-validation schemes and over external test dataset.

## Change in protein binding affinity prediction using structure descriptors for a direct comparison

In order to have a direct comparison of our sequence-based machine learning method to predict the change in binding free energy $(\Delta\Delta G)$ with structure-based methods under the same settings, we have used MutaBind-terms as structural descriptors of protein complexes taken from the state-of-the-art change in binding free energy prediction method developed by Li et al. (Li et al., 2016). For this purpose, we have used a number of regressors such as Support Vector Regressor (SVR), Random Forest Regressor (RFR), and XGBoost Regressor (XGBR) with MutaBind-terms to directly predict the change in affinity or to classify the change in affinity upon mutations as increasing ($+ive$) or decreasing ($-ive$) using trained regressor scores. Results obtained with these structural descriptors over the range of regressors through two different types of cross validation schemes called Leave One Complex Out (LOCO) and 10-fold are shown in Table 1, Table 2, and Fig. 1 (b).

For predicting the absolute change in binding free energy upon mutation using MutaBind-terms through 10-fold CV, we observed a maximum Pearson correlation coefficient of 0.81 along with 0.75, 1.07, and 0.71 as Spearman correlation coefficient, RMSE and absolute error, respectively with Support Vector Regressor (Table 1). Whereas by using LOCO CV, we observed a maximum Pearson correlation coefficient of 0.68 along with 0.62, 1.18, and 0.86 as Spearman correlation coefficient, RMSE, and absolute error, respectively with Support Vector Regressor (Table 1 and Fig. 1 (b)). The results obtained through 10-fold CV seem little exaggerated might be due to overlap of mutations from the same complex in train and test set as one complex has more than 10 mutations involved in the dataset. The 10-fold CV assumes known mutations of a complex to be

**Table 1. Predictive performance of change in protein binding affinity upon mutation across different regression models and cross validation schemes**

| Regressor | Features | Cross Validation | | | | | | | |
|---|---|---|---|---|---|---|---|---|---|
| | | 10-fold | | | | LOCO | | | |
| | | $P_r$ | $S_r$ | RMSE | AE | $P_r$ | $S_r$ | RMSE | AE |
| SVR | MutaBind-terms | 0.81 | 0.75 | 1.07 | 0.71 | 0.68 | 0.62 | 1.18 | 0.86 |
| | K-mer | **0.76** | **0.69** | **1.97** | **1.26** | **0.58** | **0.52** | **1.33** | **1.03** |
| | Blosum-62 | 0.71 | 0.63 | 2.79 | 1.87 | 0.44 | 0.42 | 3.04 | 2.11 |
| | Propy | 0.74 | 0.66 | 2.09 | 1.48 | 0.50 | 0.41 | 3.08 | 2.13 |
| RFR | MutaBind-terms | 0.74 | 0.72 | 1.10 | 0.74 | 0.62 | 0.60 | 1.23 | 0.90 |
| | K-mer | 0.74 | 0.71 | 2.01 | 1.32 | 0.56 | 0.49 | 2.92 | 2.13 |
| | Blosum-62 | 0.64 | 0.62 | 2.69 | 1.89 | 0.50 | 0.46 | 2.95 | 2.15 |
| | Propy | 0.74 | 0.68 | 2.10 | 1.48 | 0.52 | 0.48 | 2.71 | 1.96 |
| XGBR | MutaBind-terms | 0.74 | 0.72 | 1.10 | 0.75 | 0.61 | 0.59 | 1.27 | 0.90 |
| | K-mer | 0.68 | 0.62 | 2.71 | 1.88 | 0.50 | 0.46 | 3.07 | 2.22 |
| | Blosum-62 | 0.60 | 0.59 | 2.97 | 1.96 | 0.47 | 0.42 | 3.28 | 2.36 |
| | Propy | 0.73 | 0.67 | 2.15 | 1.46 | 0.46 | 0.40 | 2.93 | 2.06 |

$P_r$ (Pearson correlation coefficient), $S_r$ (Spearman correlation coefficient), RMSE (Root Mean Squared Error), AE (Absolute Error), SVR (Support Vsctor Regressor), RFR (Randon Forest Regressor), XGBR (XGBoost Regressor), LOCO (Leave One Complex Out). Bold faced values indicate best performance for each model. MutaBind-terms are taken from Li et al. (Li et al., 2016).

tested in training and does not give the real estimate of the generalization ability of a machine learning model for a protein complex with no known mutations during training as reported by Abbasi and Minhas on a slightly different problem (Abbasi and Minhas, 2016). This same trend was observed in the study performed by Li et al., in the case of CV1, CV2, and CV3 cross-validation schemes (Li et al., 2016). Here, it is important to mention that results reported by Pahari et al. in their method called SAAMBE-3D and Rodrigues et al. in their method called mCSM-

PPI2 were obtained using the same 5-fold and 10-fold cross-validation schemes, respectively (Pahari et al., 2020; Rodrigues et al., 2019). In this case, results obtained through LOCO CV look more realistic showing the real estimate of the generalization ability of a machine learning model for a protein complex with no known mutations during training since all the protein complexes in our dataset were non-redundant. Moreover, the observed Pearson correlation coefficient of 0.68 in our study using SVR corresponds to the Pearson correlation coefficient obtained by Li et. al., using a combination of multiple linear regression (MLR) and Random Forest (RF) with same features (Table 1, (Li et al., 2016)).

**Table 2. Accuracy of classification of change in protein binding affinity upon mutation into increasing or decreasing**

| Cross Validation | Features | Classfiers | | | | | |
| --- | --- | --- | --- | --- | --- | --- | --- |
| | | SVR | | RFR | | XGBR | |
| | | ROC | PR | ROC | PR | ROC | PR |
| 10-fold | MutaBind-terms | 0.81 | 0.64 | 0.82 | 0.72 | 0.82 | 0.73 |
| | K-mer | 0.77 | 0.56 | **0.79** | **0.63** | 0.76 | 0.58 |
| | Blosum | 0.73 | 0.49 | 0.75 | 0.54 | 0.75 | 0.55 |
| | Propy | 0.77 | 0.55 | 0.75 | 0.59 | 0.77 | 0.62 |
| LOCO | MutaBind-terms | 0.72 | 0.55 | 0.73 | 0.68 | 0.72 | 0.68 |
| | K-mer | 0.68 | 0.48 | **0.69** | **0.51** | 0.70 | 0.48 |
| | Blosum | 0.65 | 0.44 | 0.68 | 0.47 | 0.66 | 0.46 |
| | Propy | 0.66 | 0.44 | 0.67 | 0.47 | 0.65 | 0.43 |

ROC (Area under the ROC curve), PR (Area under the precision-recall curve), SVR (Support Vsctor Regressor), RFR (Randon Forest Regressor), XGBR (XGBoost Regressor) LOCO (Leave One Complex Out). Bold faced values indicate best performance for each model. MutaBind-terms are taken from Li et al. (Li et al., 2016).

Similarly, to classify the increase or decrease in binding free energy upon mutation using MutaBind-terms through 10-fold CV, we observed a maximum area under the ROC curve of 0.82 and under the precision-recall curve (PR) score of 0.73 with XGBR (Table 2). Whereas, by using LOCO CV, we observed a maximum area under the ROC curve of 0.72 and under the precision-recall curve (PR) score of 0.68 (Table 2). This classification task of binding free energy upon mutation, across two different cross-validation schemes, also follows the same trend as we have observed in predicting the absolute value of change in binding free energy upon mutation.

## Change in protein binding affinity prediction using sequence features

In this section, we present our results for the change in protein binding affinity prediction upon mutation using sequence information across different regression algorithms such as Support Vector Regressor (SVR), Random Forest Regressor (RFR), and XGBoost Regressor (XGBR) with sequence information to directly predict the change in affinity or to classify the change in affinity upon mutations as increasing ($+ive$) or decreasing ($-ive$) using trained regressor scores. Results obtained with sequence descriptors over the range of regressors through two different types of cross-validation schemes called Leave One Complex Out (LOCO) and 10-fold are shown in Table 1, Table 2, and Fig. 1 (a).

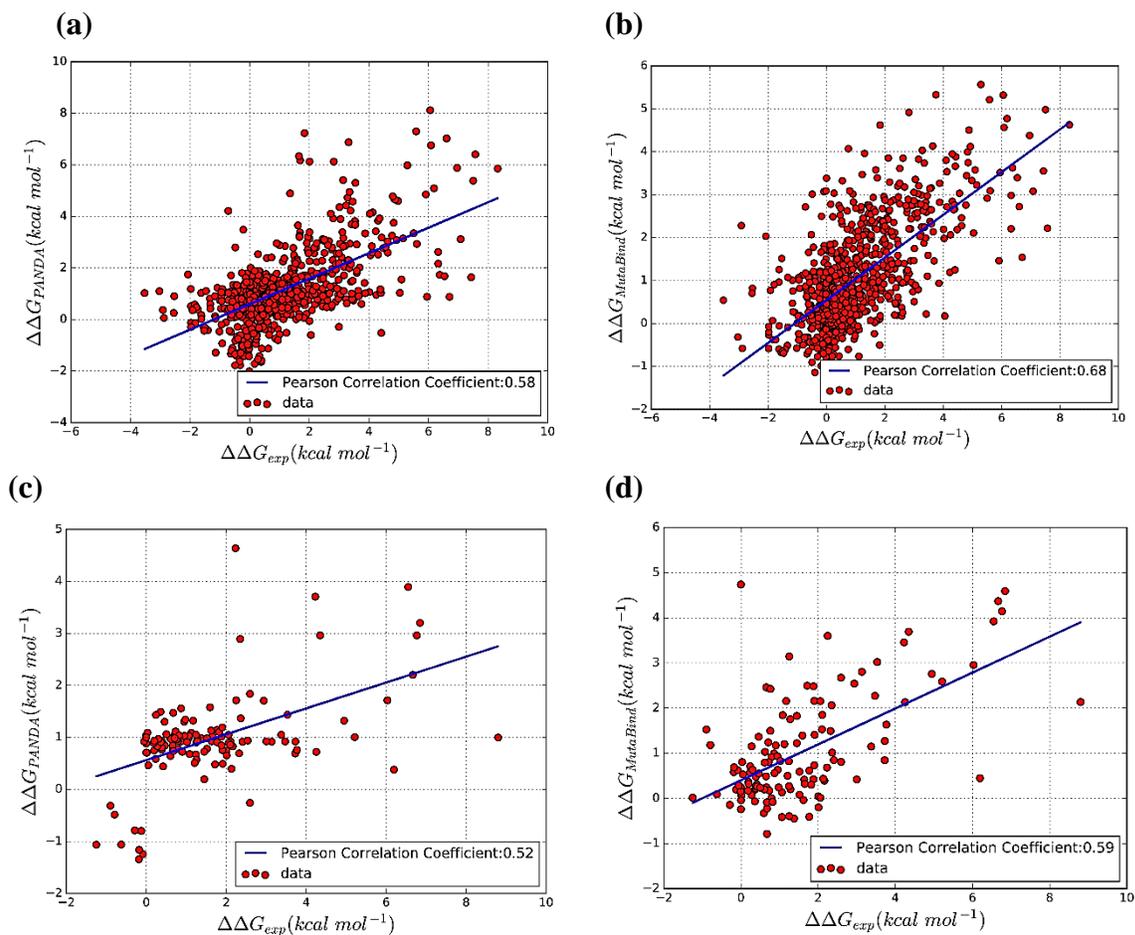

**Fig. 1. Correlation between experimental and predicted values of changes in binding affinity for all mutations.** (a) LOCO cross validation using SVR with k-mer sequence features; (b) LOCO cross validation using SVR with MutaBind structural features; (c) On validation dataset using SVR with k-mer sequence features (d) On validation dataset using SVR with MutaBind structural features

For predicting the absolute change in binding free energy upon mutation using sequence descriptors through 10-fold CV, we observed a maximum Pearson correlation coefficient of 0.76 along with 0.69, 1.97 and 1.26 as Spearman correlation coefficient, RMSE, and absolute error, respectively with Support Vector Regressor and k-mer features (Table 1). Whereas by using LOCO CV, we observed a maximum Pearson correlation coefficient of 0.58 along with 0.52, 1.33, and 1.03 as Spearman correlation coefficient, RMSE, and absolute error, respectively with Support Vector Regressor and k-mer features (Please see Table 1 and Fig. 1 (a)). As in the case of structural features in the previous section, sequence features also give exaggerated performance through 10-fold CV might be due to overlap of mutations from the same complex in train and test set as one complex has more than 10 mutations involved in the dataset. The 10-fold CV assumes known mutations of a complex to be tested in training and does not give the real estimate of the generalization ability of a machine learning model for a protein complex with no known mutations during training as reported by Abbasi and Minhas on a slightly different problem (Abbasi and Minhas, 2016). Here, it is important to mention that results reported by Chen et al., were obtained using the same 10-fold cross-validation scheme (Chen et al., 2019). In this case, results obtained through LOCO CV look more realistic showing the real estimate of the generalization ability of a machine learning model for a protein complex with no known mutations during training since all the protein complexes in our dataset were non-redundant. We have also observed that k-mer features perform consistently better across all regressors (see Table 1). This suggests that simple counts of the occurrences of individual k-mer in a sequence of a protein complex can be a good indicator of change in binding affinity upon mutations.

Similarly, to classify the increase or decrease in binding free energy upon mutation using sequence features through 10-fold CV, we observed a maximum area under the ROC curve of 0.79 and under the precision-recall curve (PR) score of 0.63 with RFR and K-mer features (see Table 2). Whereas, by using LOCO CV, we observed a maximum area under the ROC curve of 0.69 and under the precision-recall curve (PR) score of 0.51 as shown in Table 2. This classification task of binding free energy upon mutation, across two different cross-validation schemes, also follows the same trend as we have observed in predicting the absolute value of change in binding free energy upon mutation.

## Performance comparison of sequence and structure-based models on an independent external test dataset

In addition to performance comparison using LOCO and 10-fold cross-validation, we have also used an additional independent external test dataset to compare machine learning models trained using protein structural information with the proposed sequence-based approach. For this comparison, we have trained our models on the training set of 1215 mutations from 81 complexes and tested on the validation set of 129 mutations from 19 complexes. The results obtained through this process are shown in Table 3 and Fig. 1 (c) & (d).

**Table 3. Predictive performance of change in protein binding affinity upon mutation across different regression models using validation dataset**

| Features | Regressors | | | | | | | | | | | |
|---|---|---|---|---|---|---|---|---|---|---|---|---|
| | SVR | | | | RFR | | | | | | XGBR | |
| | $P_r$ | $S_r$ | RMSE | AE | $P_r$ | $S_r$ | RMSE | AE | $P_r$ | $S_r$ | RMSE | AE |
| MutaBind-terms | **0.59** | **0.48** | **1.53** | **1.07** | 0.58 | 0.51 | 1.57 | 1.10 | 0.57 | 0.45 | 1.61 | 1.09 |
| K-mer | **0.52** | **0.39** | **1.63** | **1.08** | 0.48 | 0.38 | 1.79 | 1.26 | 0.43 | 0.33 | 1.83 | 1.32 |
| Blosum-62 | 0.43 | 0.26 | 1.93 | 1.24 | 0.49 | 0.37 | 1.72 | 1.24 | 0.49 | 0.38 | 1.71 | 1.26 |
| Propy | 0.42 | 0.33 | 2.39 | 1.62 | 0.31 | 0.26 | 2.23 | 1.52 | 0.42 | 0.37 | 2.11 | 1.46 |

$P_r$ (Pearson correlation coefficient), $S_r$ (Spearman correlation coefficient), RMSE (Root Mean Squared Error), AE (Absolute Error), SVR (Support Vsctor Regressor), RFR (Randon Forest Regressor), XGBR (XGBoost Regressor). Bold faced values indicate best performance for each model. MutaBind-terms are taken from Li et al. (Li et al., 2016).

Using structural descriptors (MutaBind-terms), we obtained a maximum Pearson correlation coefficient of 0.59 along with 0.48, 1.53, and 1.07 as Spearman correlation coefficient, RMSE and absolute error, respectively with Support Vector Regressor (SVR) and k-mer features (Please see Table 3 and Fig. 1 (d)). Similarly, by using sequence descriptors (K-mer, Blosum, Propy), we obtained a maximum Pearson correlation coefficient of 0.52 along with 0.39, 1.63, and 1.08 as Spearman correlation coefficient, RMSE and absolute error, respectively. with Support Vector

Regressor (SVR) and k-mer features as shown in Table 3 and Fig. 1 (c)). These results show the comparable performance of sequence-based models to the structure-based models. These results provide further support for the use of our proposed sequence-based predictors to predict a change in protein binding affinity upon mutation where structural information is not available.

## Feature analysis for the change in binding affinity prediction upon mutations

We have used weight vectors of the best-trained models to get an insight into the role of different amino acids in predicting change in binding free energy upon mutation. Fig. 2 shows the weight vector of the trained regressor for the 1-mer features using Support Vector Regressor (SVR). We can observe that mutations with amino acids aspartic acid (D), proline (P) in the ligand (Fig. 2(a)) and glycine (G), glutamine (Q), serine (S) in receptor (Fig. 2(b)) result in an increase in binding free energy. Whereas, mutations with amino acids Glycine (G), leucine (L), threonine (T) in the ligand (Fig. 2(a)) and Cysteine (C), phenylalanine (F), methionine (M), proline (P) in receptor (Fig. 2(b)) result in a decrease in binding free energy. Most of these amino acids have already been reported as hot spots in protein interactions in different studies (Bogan and Thorn, 1998; Chakrabarti and Janin, 2002).

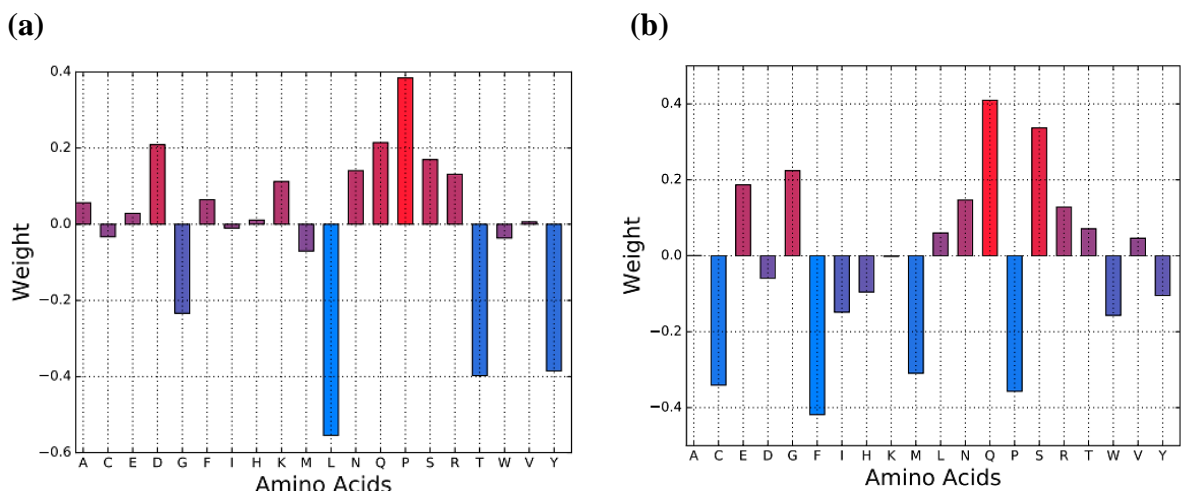

Fig. 2. Weight vectors of the trained classifiers using sequence descriptors. (a) Ligand (b) Receptor

## Conclusion

We have presented a machine learning method to predict a change in protein binding affinity upon mutations that uses protein sequence information instead of 3D structures. A comparison of the proposed method with a state-of-the-art protein sequence and structure-based change in binding affinity predictor shows that our proposed method not only shows comparable performance in cross-validation but also on an additional independent external test dataset. However, there is still a large room for improvement in solving this problem. As already suggested in recent studies, to achieve better performance in this domain, we need either a significant increase in the amount of quality affinity data or methods of leveraging data from similar problems (Abbasi et al., 2018; Dias and Kolaczkowski, 2017).

## Abbreviations

LOCO: leave one complex out, ROC: Area under the ROC curve, PR: Area under the precision-recall curve, SVR: Support Vector Regressor, RFR: Random Forest Regressor, RMSE: Root Mean Squared Error, AE: Absolute Error, CV: Cross-Validation

## Acknowledgments

WAA would like to acknowledge the support of the Higher Education Commission (HEC) of Pakistan in terms of grants: (PIN: 213-58990-2PS2-046) under indigenous Ph.D. fellowship and (PIN: IRSIP-37-Engg-02) under International Research Support Initiative Program (IRSIP).

## Availability of data and materials

All data generated or analyzed during this study are included in this paper or available at online repositories. A Python implementation of the proposed method together with a webserver is available at https://sites.google.com/view/wajidarshad/software and https://github.com/wajidarshad/panda.

## Authors' contributions

WAA conceived the idea, developed the scientific workflow, performed the experiments, analyzed and interpreted the results and was a major contributor in manuscript writing. SAA contributed to the analysis of the results and writing of the manuscript. SA helped in results interpretation and validation, and manuscript writing. All authors have read and approved the final manuscript.

## Competing interests

The authors declare that they have no competing interests.